%% file: John__Zhang__Latex__Paper__2.tex
\documentclass[12pt]{article}
\usepackage{epsfig}
\usepackage{mathtools}
\usepackage{tabulary}
\usepackage{array}
\usepackage[version=4]{mhchem}
\usepackage{float}
\usepackage[utf8x]{inputenc}
\usepackage{amsmath}
\usepackage{caption}

\input econfmacros.tex
\def\Title#1{\begin{center} {\Large {\bf #1} } \end{center}}

\begin{document}

\Title{Studies of Silver Bromide Clusters Isotopic Properties and Their Applications}

\bigskip\bigskip


\begin{raggedright}  

{\it John Hongguang Zhang\index{Zhang, J.}\\
Altamont Research\\
Altamont, NY12009}
\bigskip\bigskip
\end{raggedright}

{\bf Abstract:}  The quantum size effect (QSE) studies of metallic and semiconductor nanoparticles have received considerable attention for their applications in superconductivity, visible light emission, quantum dot devices and industrial catalysts. The QSE properties of semiconductor AgBr nanoparticles have been thoroughly characterized for cluster larger than three nanometers in diameter, the QSE studies of clusters in the quantum size regime less than 3nm have been historically difficult to describe owing to weak optical absorption and challenge in keep cluster in small size range against self aggregation. Such difficulties have precluded probing the properties of these small clusters in many natural and engineered processes. Our previous investigation of silver bromide ionic clusters that prepared via the electroporation of vesicles using direct laser desorption -time of flight- mass spectrometry (DLD-TOF-MS) had successfully revealed relation ship between  cluster structure and the UV absorption band shift. The turn-around point had been experimentally identified to be $ (Ag_3Br_2)^+ $, which is close to the theoretical prediction. Despite those successes, there a lot of questions not be answered.  These unanswered questions include why DLD-TOF-MS spectra show an unsymmetrical cluster finger peaks? why the large clusters have much lower DLD-TOF-MS spectra intensity? Why the cluster UV absorption have such wide band and it take so long time to observe the band shift? In this paper, we reexamine our previous theoretical work and systematically  answer the above questions with symmetry and probability principles in molecular cluster growth range. We also discussed the isotopic properties of silver bromide clusters and their possible applications on the dark matter detection based on recent findings that isotopic atoms can decay while interact with dark matters.

\section{Introduction}

A most striking property of metal and semiconductor nanocrystals is the massive changes in physical properties as a function of the size ~\cite{Rossetti}. This is usually called Quantum Size Effects (QSE)~\cite{Onushchenko}. It appears that QSE have been used unknowingly for years, even centuries, to produce the aesthetic beauty of "stained" glass. Nanomaterials exhibit physical and chemical properties very different from those of their bulk counterparts, often resulting from enhanced surface interactions or quantum confinement~\cite{Scholl} . In the past decades, QSE had found wide range applications including superconductivity ~\cite{Guo}, visible light emission~\cite{Cullis} , quantum dot photodetectors~\cite{Konstantatos} , and industrial catalysts~\cite{Behrens} .  QSE study is also very significant to fundamental science since the energy structure in the molecular and atomic stage is well defined by quantum mechanics and in the bulk crystal stage by band theory. However in the intermediate stage, the energy structure has not yet been well modeled~\cite{Zhang2019A}. There are many open questions in this stage and new physics and chemistry are sure to be discovered in the future ~\cite{Alivisatos,Alivisatos1,Chen1997A} .

  The growth of silver halide particles has been the subject of many studies in conjunction with their application in photographic processes\cite{James} and recently molecular devices\cite{Zhang2017A, Zhang2019B}. The growth mechanism is particularly important in such a process, since the sensitivity of the photographic material depends on its grain size as well as its crystal structure~\cite{James,Tadaaki,Ehrlich}. The QSE properties of AgBr nanoparticles have been studied intensively in several ways so far. They are size dependence of exciton energies~\cite{Chen1994A,Masumoto}, size dependence of exciton lifetime~\cite{Johannson,Marchetti,Kanzaki, Freedhoff}, size dependence of Raman scattering cross section~\cite{von,Scholle2,Vogelsang}     and the size dependence of the UV absorption spectra ~\cite{Tanaka,He,Zhang1997A}.  Although the QSE of nano AgBr semiconductor quantum dots with diameters exceeding 3 nm have been well characterized ~\cite{Brus,Chen1994B}, QSEs of AgBr nanoparticles below this size are poor understood. Experimentally synthesize the clusters in this smaller size regime has been extremely challenging since the spontaneous self-aggregation of the constituent molecules must be controlled and halted at the quantum size level~\cite{Zhang2000A}  .  Furthermore, optical detection is hampered by these particles’ diminishing scattering and absorption intensities. At this small particle size range, absorption measurements can give information without serious disturbance caused by light scattering, but the challenge is to detect the weak absorption of these small particles~\cite{Tanaka1} . For example, pulse radiolysis was used to study the formation of the AgBr monomer and the $(AgBr)_2$ dimer in solution. $AgBr$  monomer absorption at 295 nm was successfully observed shift to blue to form dimer $(AgBr)_2$  at 285 nm ~\cite{Zhang1997A}. However, the absorption intensity decreasing and   the absorption peak broaden made it difficult to observe the trimer and tetramer with this method and we don't know why.  Method of preparation of AgBr quantum dots via electroporation of vesicles let us successfully synthesized the $AgBr$  clusters from 5Å to 2nm.  For the first time, we observed the entire blue-shift (274nm to 269nm) followed by red-shift (269nm to 273nm) of the absorption band that is associated with the growth of the silver bromide clusters. The turn-around point is at 269 nm ~\cite{Zhang2000A}. However we don't know why the cluster UV absorption have such wide band and it takes so long time to observe the band shift? The molecular and electronic structures and UV absorption spectra of neutral $(AgBr)_n$ clusters (n =1-9) were investigated in both the gas phase and in a dielectric medium by Ab initio density functional theory (DFT). Our computational results parallel the experimental trends and predict that the maximum blue shift occurs at the trimer or the tetramer~\cite{Zhang2000B} . By using direct laser desorption mass spectra, we successfully observed $(Ag_2Br)^+$ , $(Ag_3Br_2)^+$ , and $(Ag_4Br_3)^+$ clusters in the same three samples that were used in the UV absorption experiments. For the first time, we experimentally confirm that the turn-around point cluster structure is $(Ag_3Br_2)^+$. This is very close to our theoretical prediction ~\cite{Zhang2000C,Zhang2012A}. However questions, such that why DLD-TOF-MS spectra show an unsymmetrical cluster finger peaks? why the large clusters have much lower DLD-TOF-MS spectra intensity? are not answered.  
In this paper, we reexamine our previous theoretical work and systematically  answer the above questions. We also discussed the possible applications of the  isotopic properties of silver bromide cluster based on our studies.

\section{Isotopic Properties Studies of Silver Bromide Clusters }
\subsection{Reexamination of Silver Bromide Cluster Growth Theories }
In molecular and small cluster size range, several growth mechanism of silver bromide clusters have been proposed. By using stopped-flow method, where the concentration of the bromide ions is much larger than the concentration of silver ions in the solution, both Tanaka~\cite{Tanaka1985A} and Ehrlich  ~\cite{Ehrlich} proposed the following polybromoargentate mechanism:\\
\begin{flushleft}
\ce{Ag+ + Br- ->AgBr}\\
\ce{AgBr + iBr- ->AgBr_{(i-1)}^{i-}}\\
\ce{nAgBr_{(i-1)}^{i-} -> Ag_nBr_{(n+j)}^{j-} + (ni-j)Br-}\\
\end{flushleft}
Here, n is the number of silver-bromide ion pairs needed for the formation of a primary particle (nucleus) of silver bromide, and was estimated to be 4. Symbol i represents the number of excess bromide ions present in the transient complex. Symbol j has been used to include the possibility that the particles are negatively charged due to adsorbed or bonded excess bromide ions (polynuclear complex anions). This mechanism is supported by the observation of the 230nm UV absorption band which was assigned to be the transient polybromoargentate complex anions $AgBr_i ^{(i-1)-} (i=1,2,3,4)$  in both of their experiments. \\

By using the pulse radiolysis method ~\cite{Zhang1997A}, where the concentration of the bromide ions produced in the solution is almost the same as the concentration of silver ions in the solution, we had proposed the following growth mechanism:\\
\begin{flushleft}
\ce{Ag+ + Br- ->AgBr}\\
\ce{(AgBr)_{n-1} + Ag+ -> (Ag_nBr_{n-1})+}\\
\ce{(Ag_nBr_{n-1})+ + Br- -> (AgBr)_n}\\
\end{flushleft}
We reported the observation of AgBr monomer with the UV absorption band at 295nm and the AgBr dimer with the UV absorption band at 285nm ~\cite{Zhang1997A}. \\

Our experiments that prepared the AgBr clusters via the electroporation of vesicles let us  observed the entire blue-shift (274nm to 269nm) followed by red-shift (269nm to 273nm) of the absorption band that is associated with the growth of the silver bromide clusters. In this experiment, the following  growth mechanism was proposed\cite{Zhang2000A}:\\

\begin{figure}[H]
\begin{center}
\epsfig{file=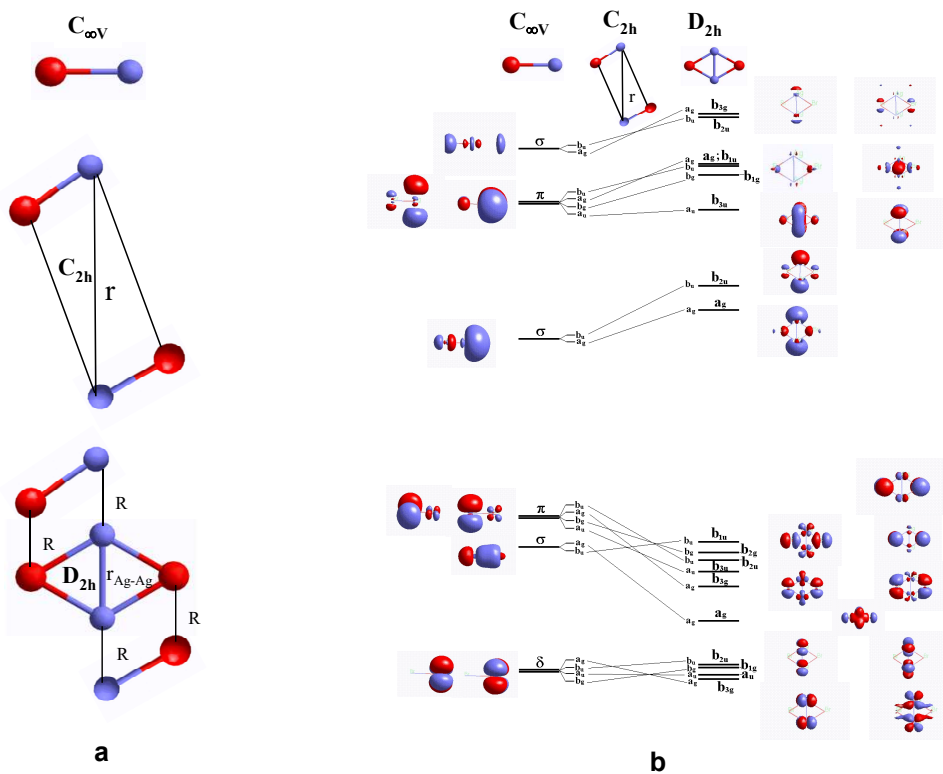,height=4.8in}
\caption{Formation of the dimer from two monomers with overall $C_{2h}$ symmetry and molecular orbital energy level correlation diagram for dimerization.
{\bf a}, Illustrates how the dimer can be constructed from two monomers by varying only one geometrical parameter R.  {\bf b},  Shows the resulting correlation diagram,  the monomer ($C_{\infty V}$) and the dimer ($D_{2h}$) share a common symmetry of $C_{2h}$. The results had been used to explain the molecular origin of the absorption blue shift from monomer to dimer}
\label{fig:figure1}
\end{center}
\end{figure}

\begin{flushleft}
\ce{Ag+ + Br- ->AgBr}\\
\ce{n(AgBr) -> (AgBr)_n}\\
\end{flushleft} 
 To understand those growth mechanisms that proposed in different experiment conditions and their UV absorption spectra behavior, we present the results of our theoretical studies on both neutral silver bromide clusters ($(AgBr)_n, n=1-9$)~\cite{Zhang2000B} and silver bromide ion clusters($ (Ag_nBr_{n-1})^+  n=2-4$) ~\cite{Zhang2000C, Zhang2012A} and their UV absorption spectra. HOMO-LUMO energy gaps,
singlet-triplet energies, and CIS calculations demonstrate first
a blue and then a red shift of the absorption band upon cluster
growth which parallel the experimental trends. The molecular
origin of the blue/red shift can be explained by use of molecular
orbital energy level correlation diagrams.  Figure~\ref{fig:figure1} illustrates how the dimer can be constructed from two monomers by varying only one geometrical parameter, and the resulting molecular
orbital energy level correlation diagram.    
For the dimerization, the monomer($C_{\infty V}$)  and the dimer ($D_{2h}$) share a common symmetry of $C_{2h}$. The HOMOs of the monomers are of $ \pi $ symmetry (essentially 4p orbitals on Br) which transform as $ b_u + a_g + b_g + a_u $  in $C_{2h}$ symmetry. The next lowest orbitals are of $\sigma$  symmetry $(a_g + b_u)$  in   $C_{2h}$ and are dominantly bromine $4p_{\sigma}$ and silver $4_{{dz}^2} $(the bromine $4s$ orbital is lower in energy that the silver $3d$ orbitals, and may to a first approximation be neglected). The next lowest orbitals are the$Ag$  4d orbitals of $\delta$ symmetry $ (b_u + a_g + b_g + a_u )$ in $C_{2h}$ . The $\pi$ type d orbitals are of lower energy and are not shown on the diagram. The LUMOs are predominately $ 5s/5p_{\sigma} $ silver atomic orbitals which transform as $a_g + b_g$ in $C_{2h}$. The second LUMOs are vacant 5p orbitals on Ag which transform as $ b_u + a_g + b_g + a_u $  in  $C_{2h}$. Thus, the first six HOMOs and the first six LUMOs transform as $  2a_g + a_u +b_g +2b_u $ in $C_{2h}$. It is the mixing of these orbitals which results in a stabilization of the occupied orbitals and a destabilization of the unoccupied orbitals. The net effect is an increase in the HOMO-LUMO gap, as illustrated in Figure~\ref{fig:figure1}.  Note that only one of the occupied orbitals of the dimer (the $a_g$ orbital arising from mixing of the two $\sigma$ orbitals of the monomer) has significant Ag-Ag bonding character. This is a general trend also seen in the higher clusters: The short Ag-Ag interactions do not arise from strong Ag-Ag bonding, but rather from the fact that formation of silver clusters allows for the efficient overlap of bromine-occupied atomic orbitals with the unoccupied orbitals of several silver atoms. 

Indeed, the Ag-Ag Mulliken overlap ~\cite{Mulliken, Holger},  population at the B3P86/SB level is actually negative, and a natural bond order analysis ~\cite{Foster, Reed}, at the same level describes the dimer as having four Ag-Br bonds and no Ag-Ag bond.  Silver clusters formation inside AgBr clusters had been detected in our early experiments of surface enhanced Raman scattering of pyridine in AgBr Sol~\cite{Zhang1991A} . It was found that $Ag_4 $ cluster gives the largest contribution to the SERS intensity \cite{Zhang1995A} and the intensity change can be used to study the nano AgBr cluster fractal aggregation structures\cite{Zhang1994A} .

\begin{figure}[H]
\begin{center}
\epsfig{file=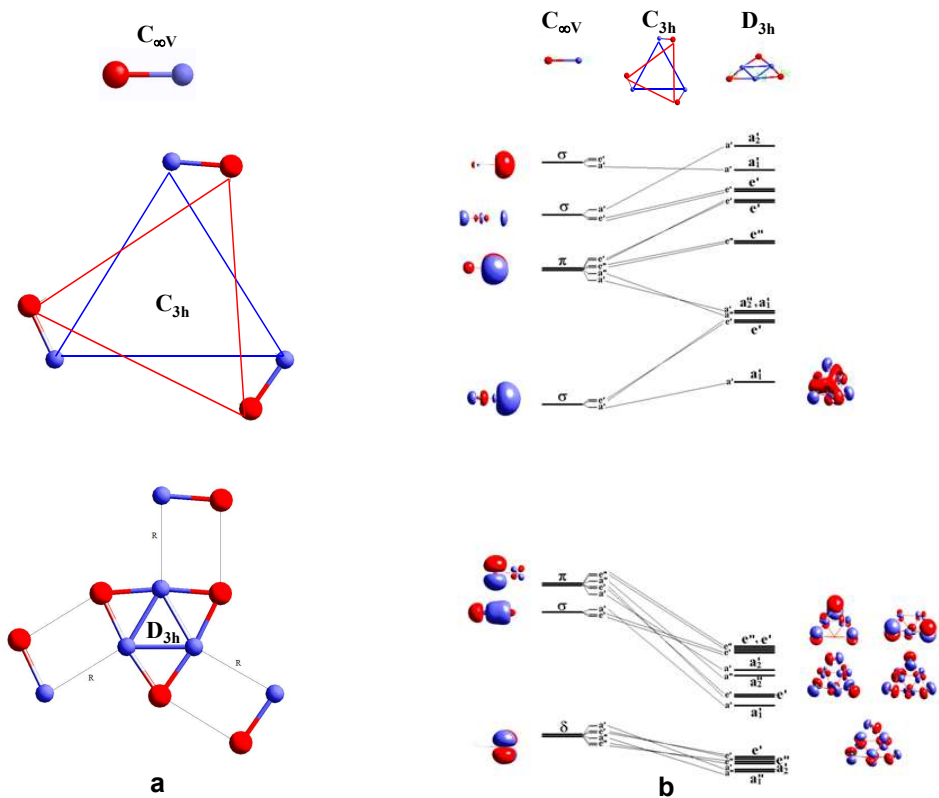,height=4.8in}

\caption{Formation of the trimer from three monomers with overall $C_{3h}$ symmetry and molecular orbital energy level correlation diagram for trimerization.
{\bf a}, Illustrates how the trimer can be constructed from three monomers by varying only one geometrical parameter R.  {\bf b},  Shows the resulting correlation diagram,  the monomer ($C_{\infty V}$) and the trimer ($D_{3h}$) share a common symmetry of $C_{3h}$. The results had been used to explain the molecular origin of the absorption blue/red shifts from monomer to trimer.}
\label{fig:figure2}
\end{center}
\end{figure}

\begin{figure}[H]
\begin{center}
\epsfig{file=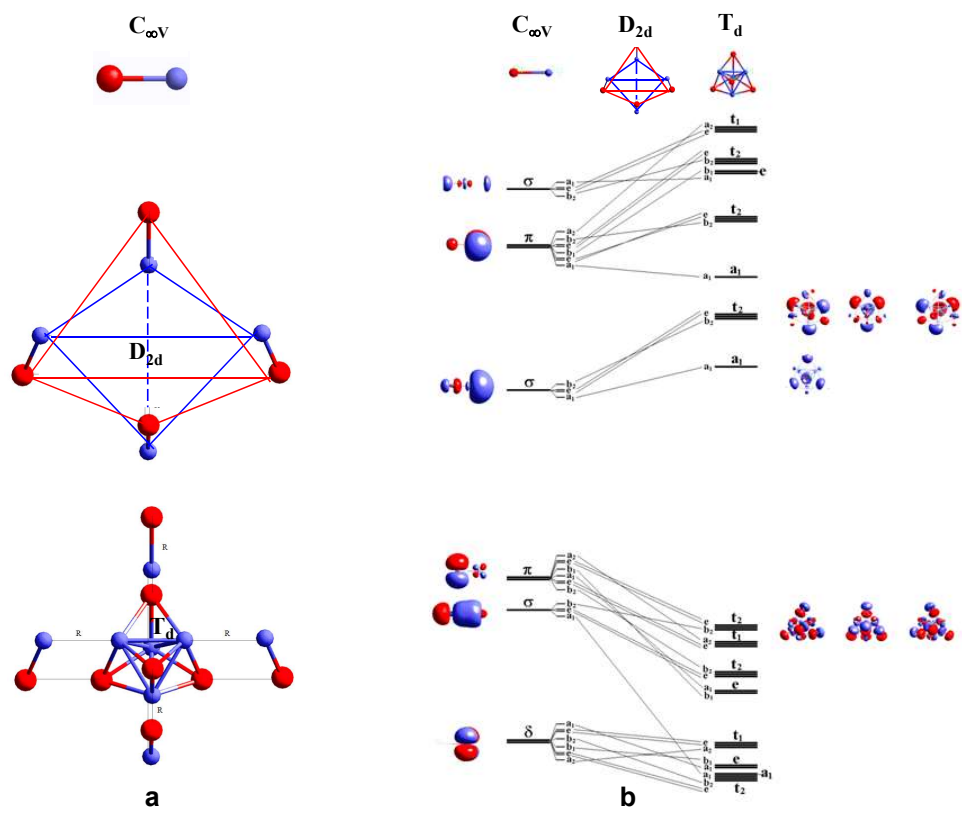,height=4.8in}
\caption{Formation of the tetramer from four monomers with overall $D_{2d}$ symmetry and molecular orbital energy level correlation diagram for tetramerization.
{\bf a}, Illustrates how the tetramer can be constructed from four monomers by varying only one geometrical parameter R. {\bf b},  Shows the resulting correlation diagram,  the monomer ($C_{\infty V}$) and the tetramer ($T_d$) share a common symmetry of $D_{2d}$. The results had been used to explain the molecular origin of the absorption blue/red shifts from monomer to tetramer.
}
\label{fig:figure3}
\end{center}
\end{figure}

The same process was used to obtain the correlation diagrams for trimerization and tetramerization. For trimerization, Figure ~\ref{fig:figure2} , the common symmetry is  $C_{3h}$  (the symmetry of the trimer is $D_{3h}$), while for tetramerization , Figure ~\ref{fig:figure3} , the common symmetry is $D_{2d}$ (the symmetry of the tetramer is $T_d$). Upon proceeding from the dimer to the trimer, the number of Ag-Br interactions increases from four to six. This allows for increased orbital overlap between the HOMOs and LUMOs of the monomeric species and results in a larger HOMO-LUMO gap.\\
   
   One might expect the same phenomena to occur for the tetramer, where the number of Ag-Br interactions increases to 12, but the HOMO-LUMO gap actually decreases slightly in the gas phase, and remains about the same for the calculations done in a dielectric medium ~\cite{Zhang2000B,Zhang2000C}. As in the case of the dimer, Ag-Ag interactions are not the dominant feature in these clusters. Ag- Ag Mulliken overlap populations are again negative. As one proceeds to larger clusters, there are no longer enough bromine atoms to saturate all of the faces of the tetrahedra. This produces low-lying LUMOs localized on the empty faces of $Ag_4 $ tetrahedra and lowers the HOMO-LUMO gap. Our computational results parallel the experimental trends and predict that the maximum blue shift occurs at the trimer or the tetramer~\cite{Zhang2000B} . By using direct laser desorption mass spectra, we successfully observed $(Ag_2Br)^+, (Ag_3Br2)^+ and (Ag_4Br3)^+ $ clusters in the same three samples that were used in the UV absorption experiments. For the first time, we experimentally confirm that the turn-around point cluster structure is $(Ag_3Br_2)^+$. This is very close to our theoretical prediction ~\cite{Zhang2000C,Zhang2012A}. However questions, Why in our pulse radiolysis experiments, it take only 400ns for us to observe the monomer and 1.5 $\mu$ s to observe dimer blue shift and we can not observe trimer and tetrama signal~\cite{Zhang1997A} while in our electroporation of vesicles experiments the cluster UV absorption have such wide band and it take so long time to observe the band shift (5 hours for the blue shift and 6 hours for the red shift~\cite{Zhang2000A}) ? why DLD-TOF-MS spectra show an unsymmetrical cluster finger peaks? why the large clusters have much lower DLD-TOF-MS spectra intensity~\cite{Zhang2000C,Zhang2012A} are not answered. In the following two sections, we will systematically answer these questions through the introduction of symmetry and probability principle in molecular cluster growth range based on the theoretical and experimental studies we examined in this section

\subsection{Symmetry and Probability Principle in Molecular Cluster Range }   

In order to systematically answer the questions we have during our serious experiments, we proposed the following symmetry and probability principle in molecular cluster growth range :\\

\begin{flushleft}
1) The less elements , the higher probability to form the cluster.\\
2) The higher symmetry and more elements clusters have lower probability to form comparing to lower symmetry and less elements clusters in the same environments.\\
3) The high symmetry more elements clusters have a trend to form high HOMO LUMO energy gap and have high stability.\\
\end{flushleft}

\begin{figure}[H]
\begin{center}
\epsfig{file=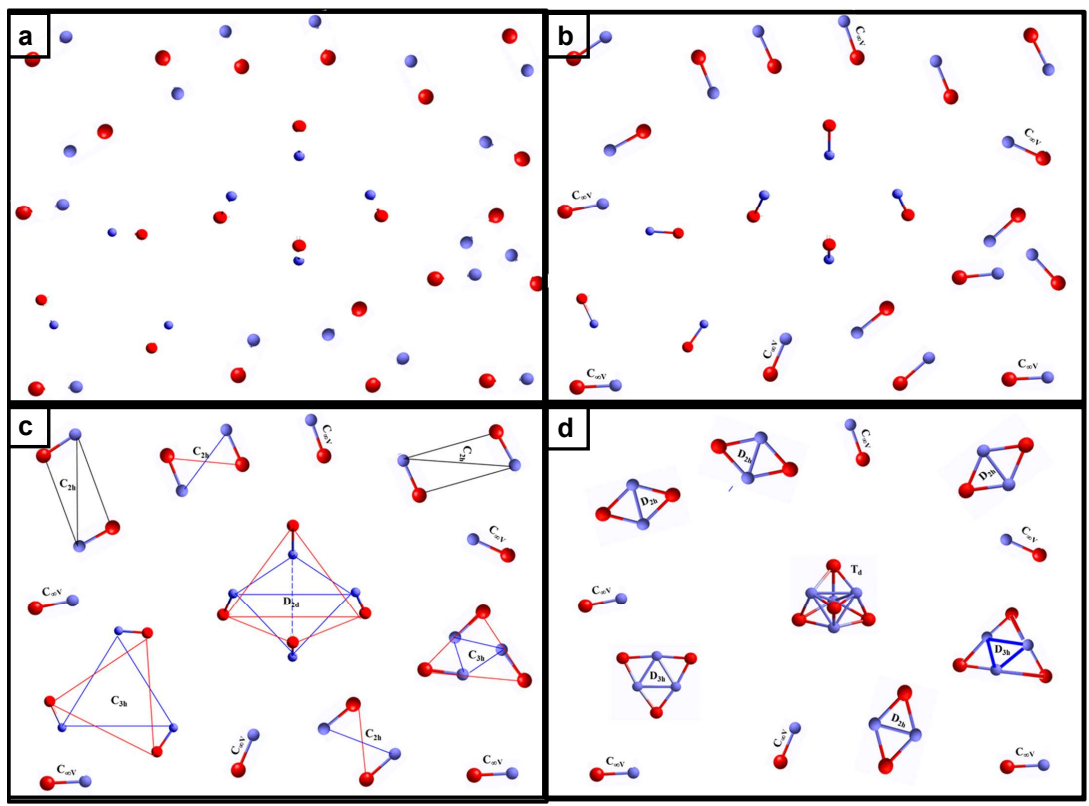,height=4.8in}
\caption{Symmetry and probability principle for molecular range cluster formation. {\bf a}, Random and equal amount of silver (blue) and bromide (red) ions distributed in solvent or vacuum environment at t=0. {\bf b}, Form monomer quickly. {\bf c}, Trend to form dimer, trimer and tetramer. {\bf d}, Form dimer, trimer and tetramer }
\label{fig:figure4}
\end{center}
\end{figure}

We can use Figure ~\ref{fig:figure4} to understand what we stated in the principle. Figure ~\ref{fig:figure4}{\bf a} describe the random distributed equal amount of silver (blue) and bromide (red) ions in solvent or vacuum environment at t=0. Based on the symmetry and probability principle 1, we know that the most easy formed cluster should be monomer, this had be described in Figure ~\ref{fig:figure4}{\bf b}.  

In our pulse radiolysis experiment environment, this took only 400ns \cite{Zhang1997A}. Next step is monomers form the dimer, trimer and tetramer etc, based on the symmetry and probability principle 2, the dimer will have much high probability to form comparing to trimer and tetramer, this trend has been shown in  Figure ~\ref{fig:figure4}{\bf c}. Finally  dimer, trimer and tetramer etc can be formed simultaneously based on the monomers symmetry distribution as shown in Figure ~\ref{fig:figure4}{\bf d}. In our pulse radiolysis experiment environment, this took 1.5 $\mu$ s and our experiments can not observe trimer and tetrama signal~\cite{Zhang1997A}. This can easily be understood by the the symmetry and probability principle 2 since the probability to form the trimer and tetramer is too small to be observed. 

\begin{figure}[H]
\begin{center}
\epsfig{file=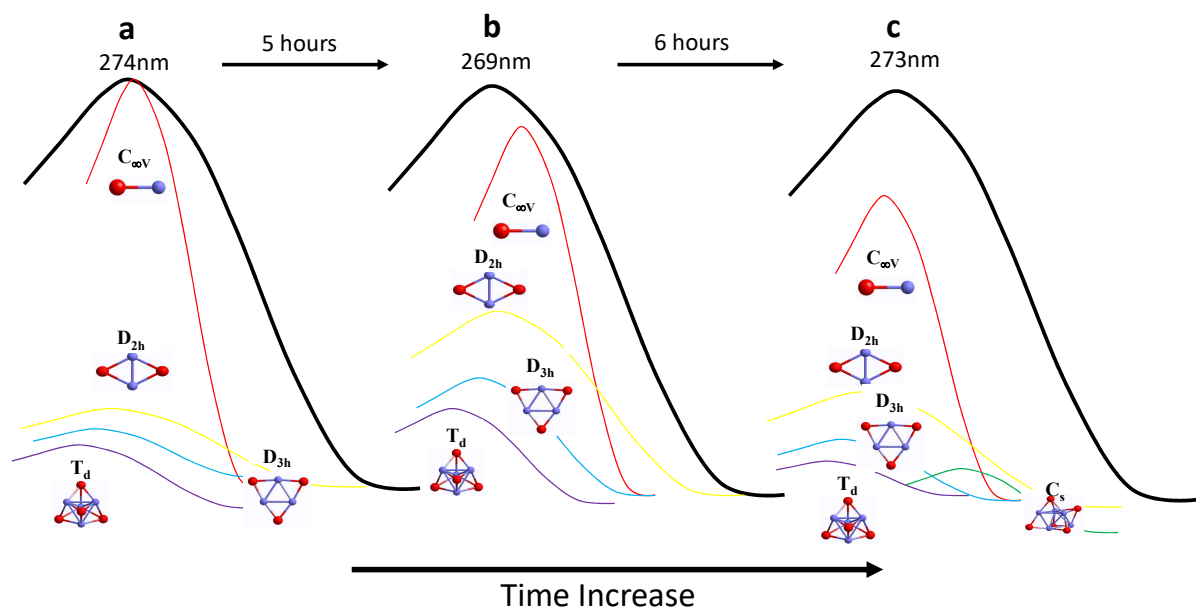,height=4.8in}
\caption{UV absorption spectra prepared via the electroporation of vesicles show wide absorption bands and long time for observation of band shift }
\label{fig:figure5}
\end{center}
\end{figure}

Our electroporation of vesicles experiments can also be understood by the symmetry and probability principle, this has been shown in  Figure ~\ref{fig:figure5}. 
Figure ~\ref{fig:figure5}{\bf a} show the wide absorption band right after the electroporation of vesicles, this corresponds to the 1.5 $\mu$s step of our pulse radiolysis experiment, or the Figure ~\ref{fig:figure4}{\bf d} step, the wide absorption bands actually is the composition of dimer, trimer, tetramer etc. In this step,  dimer is dominate absorption spectra. After Figure ~\ref{fig:figure4}{\bf d} step, the dominated dimers will repeat Figure ~\ref{fig:figure4}{\bf b} to  Figure ~\ref{fig:figure4}{\bf d} cycle to form more and more trimers and tetramers.  Figure ~\ref{fig:figure5}{\bf b} show the results that after numerous Figure ~\ref{fig:figure4}{\bf b} to  Figure ~\ref{fig:figure4}{\bf d} cycles, the trimer and tetramer amounts reach the maximum in the solution.  This cause the wide band shift to blue and is also the reason why it takes so long time (5 hours) to shift. After trimers and tetramers reach the maximum in the solution, they will continue a similar growth cycle with rest monomers to form higher clusters such as hepatamers, hexamers,...etc. This  cause the wide band shift to red and is also the reason why it takes so long time (6 hours) to shift as shown in Figure ~\ref{fig:figure5}{\bf c}. Another reason for the long time absorption band shift can be the interaction between the formed clusters and the surface of the vesicles. This had been discussed in our previous studies\cite{Zhang2000C} .

\subsection{Isotopic Properties Studies of Silver Bromide Clusters}

\begin{figure}[H]
\begin{center}
\epsfig{file=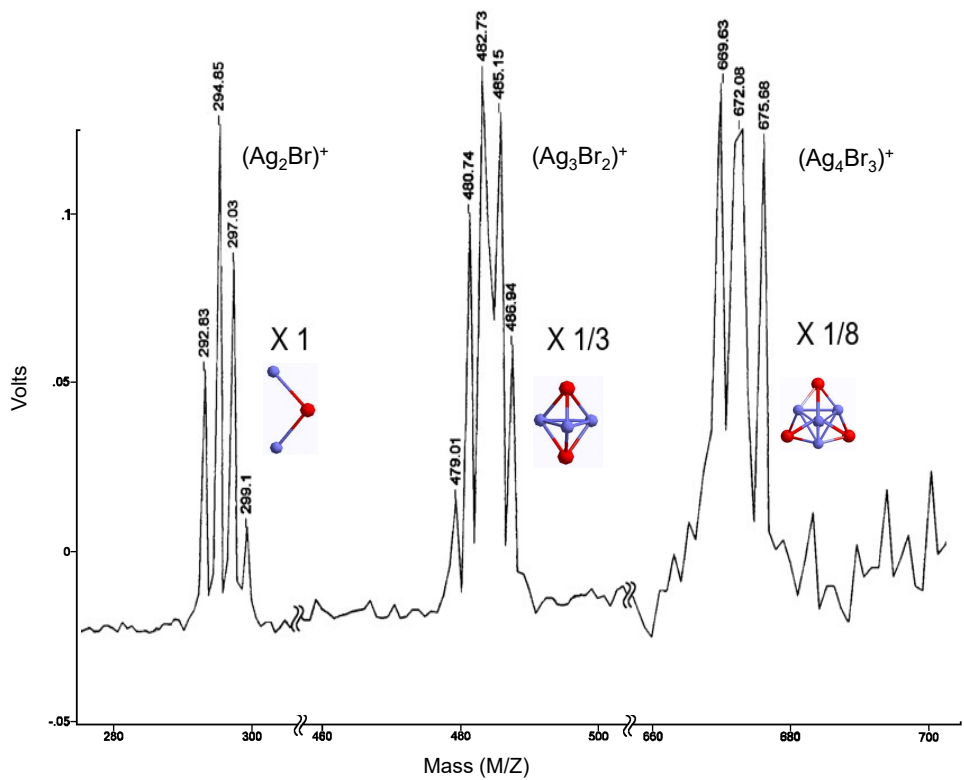,height=4.8in}
\caption{DLD mass spectra of AgBr clusters showing the $(Ag_2Br)^+$ , $(Ag_3Br_2)^+$ , and $(Ag_4Br_3)^+$ species}
\label{fig:figure6}
\end{center}
\end{figure}

Figure  ~\ref{fig:figure6} presents DLD-TOF-MS spectra of three samples that are same as we used in the UV absorption experiments prepared via the electroporation of vesicles (~\cite{Zhang2000C, Zhang2012A}).Since in this work, our DLD-TOF-MS experiments with the same samples used for the UV absorption experiments had successfully observed the $(Ag_nBr_{n-1})^+ n=2,3,4 $ clusters as described in the previous section. Although our current experimental setup does not permit direct detection of the initially formed AgBr monomer and dimer (AgBr)2 (with absorption bands at 295 and 285 nm, respectively,~\cite{Zhang1997A} ),  we do observe the entire blue-shift followed by red shift of the absorption band (274 nm (5 h) →269 nm (6 h) →273 nm, ~\cite{Zhang2000A}) associated with the slow growth of the clusters through the molecular size to the typical quantum size regime. Density functional calculations at the B3P86/LB level suggest that the turn-around point (269 nm) in the band shift occurs at the trimer $(AgBr)_3 $ or tetramer $(AgBr)_4 $ stages of cluster growth, at which point the particle size is ˜5 Å (~\cite{Zhang2000A}).  Here, with DLD-TOF-MS, we experimentally observed the turn-around point cluster structure is $ (Ag_3Br_2)^+$. This is very close to the theoretical prediction. In Figure  ~\ref{fig:figure6}, an unsymmetrical cluster finger peaks was shown and also the large clusters have much lower DLD-TOF-MS spectra intensity. What's the mechanism behind them? This can be answered through the studies of isotopic properties of silver and bromide. 
Naturally occurring silver (Ag) is composed of two stable isotopes ${}^{107}Ag$ and ${}^{109}Ag$,  ${}^{107}Ag$ has 51.839\% natural abundance and 106.905097 u isotopic mass.  ${}^{109}Ag$  has 48.161\% natural abundance and 108.904752u isotopic mass. Naturally occurring bromine (Br) is also composed of two stable isotopes ${}^{79}Br$ and ${}^{81}Br$. ${}^{79}Br$ has 50.69\% natural abundance and 78.9183371 u isotopic mass.${}^{81}Br$ has 49.31\% natural abundance and 80.9162906u isotopic mass.  By considering these isotopic effects, all the peaks observed in the experiment can be assigned exactly within the error range of the experiment. The assignments have been shown in Table ~\ref{tab:Iso}.  The sample with UV absorption blue-shift start point (274nm) observed 4 peaks and can be assigned to be the $(Ag_2Br)^+$ ionic cluster based on the isotopic calculation formula.  The sample with UV absorption turn-around point (269nm) observed 5 peaks and can be assigned to be the $ (Ag_3Br_2)^+ $ ionic cluster with one peak missing. The sample with UV absorption red-shift end point (273 nm) observed 3 peaks and can be assigned to be the $(Ag_4Br_3)^+ $ ionic clusters with five peaks missing. The unsymmetrical cluster finger peaks can be easily understood through the compare of the abundance of Ag and Br isotopic atoms. For example, in Table~\ref{tab:Iso}, $(Ag_2Br)^+$ cluster peaks distribution is 1:2:2:1, the left 1 peak contribute by cluster ${}^{107}Ag_2{}^{79}Br_2$, the right 1 peak contribute by cluster ${}^{109}Ag_2{}^{81}Br_2$, since both ${}^{107}Ag$ and ${}^{79}Br$ have higher natural  abundance than ${}^{109}Ag$ and ${}^{81}Br$, this is why left 1 peak has much high intensity compare to right 1 peak and give the unsymmetrical characterization. Through the calculation of the total abundance of the two clusters, we can give the quantitative ratio between these two peaks.  

\begin{figure}[htb]
\begin{center}
\epsfig{file=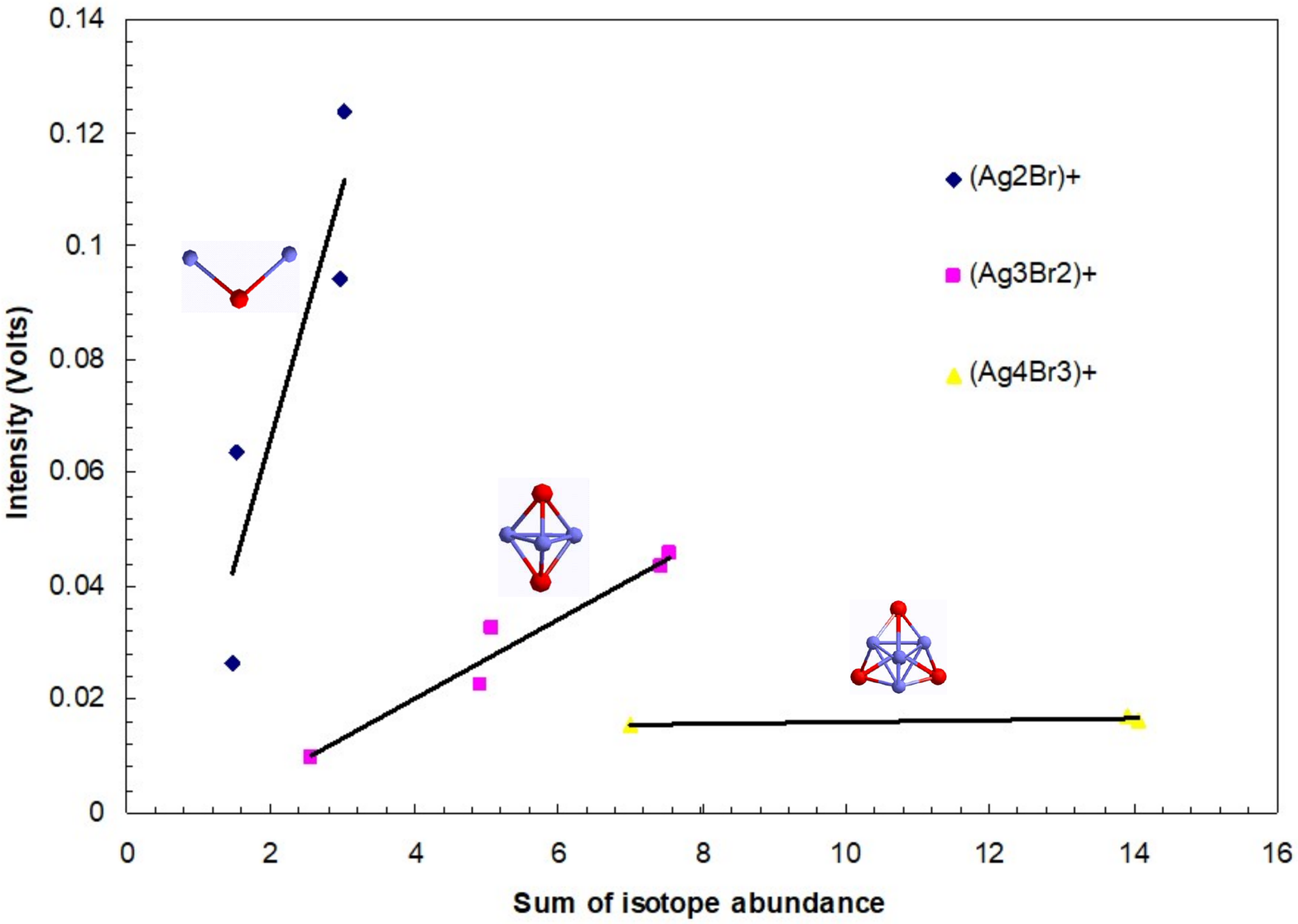,height=4.8in}
\caption{DLD mass spectra intensity vs sum of abundance for the $(Ag_2Br)^+$ , $(Ag_3Br_2)^+$ , and $(Ag_4Br_3)^+$ species}
\label{fig:figure7}
\end{center}
\end{figure}

  The other thing we observed in this experiment is the DLD-TOF-MS intensity decrease as the ionic cluster size increase.  Nagao et al studied the mass spectra of $Ag_xBr^+ $ x=2,4,6,8,10 ~\cite{Nagao} . They also observed the mass spectra intensity decrease as the cluster size increase . Hermite et al studied the abundance spectra of silver bromide clusters ~\cite{Hermite}. But non of them explained why. Our previous work has studied this trend in detail ~\cite{Zhang2012A}. The characterization had been summarized in Figure ~\ref{fig:figure7} and Equation~\ref{equ:a}

\begin{equation}
 I_n=\alpha_n \times\Sigma_a+\delta_n
\label{equ:a}
\end{equation}

Here $I_n$is the peak intensity of $(Ag_nBr_{(n-1)})^+$ ionic cluster.  $\alpha_n$ and $\delta_n$ are constant for fixed structure $(Ag_nBr_{(n-1)}^+)$  ionic cluster. The sum of the isotopic abundance of each peak $\Sigma_a$ is calculated based on the isotopic formula listed in Table ~\ref{tab:Iso}. For example, for the $(Ag_2Br)^+$  cluster , the peak with isotopic formula  ${}^{107}Ag_2{}^{79}Br_2$  can be calculated as: 

\begin{equation}
\Sigma_a =2\times0.51839+0.5069= 1.54368
\label{equ:b}
\end{equation}

\begin{table}[p]
\begin{center}

\begin{tabular}{p{1.6cm}p{1.8cm}p{2cm}p{1.0cm}p{3.8cm}p{2cm}} \hline 
MS exp value & Exp intensity &Calculated value & Merge peak & Isotopic formula & Cluster type\\ \hline 
292.83 &  0.064   & 292.728   & 1 & \ce{^{107}Ag_2}\ce{^{79}Br}& $ (Ag_2Br)^+$\\
294.85 &  0.124   & 294.726   & 2 & \ce{^{107}Ag_2}\ce{^{81}Br}& $ (Ag_2Br)^+$\\
 &    & 294.728   &  & \ce{^{107}Ag} \ce{^{109}Ag}\ce{^{79}Br}& $ (Ag_2Br)^+$\\
297.03 &  0.094   & 296.728   & 2 & \ce{^{109}Ag_2}\ce{^{79}Br}& $ (Ag_2Br)^+$\\
 &    & 296.726   &  & \ce{^{107}Ag} \ce{^{109}Ag}\ce{^{81}Br}& $ (Ag_2Br)^+$\\
299.10 &  0.026   & 298.726   & 1 & \ce{^{109}Ag_2}\ce{^{81}Br}& $ (Ag_2Br)^+$\\
479.01 &  0.010   & 478.552   & 1 & \ce{^{107}Ag_3}\ce{^{79}Br_2}& $ (Ag_3Br_2)^+$\\
480.74 &  0.032   & 480.550   & 2 & \ce{^{107}Ag_2}\ce{^{79}Br}\ce{^{81}Br}& $ (Ag_3Br_2)^+$\\
 &    & 480.552   &  & \ce{^{107}Ag_2}\ce{^{109}Ag}\ce{^{79}Br_2}& $ (Ag_3Br_2)^+$\\
482.73 &  0.046   & 482.548   & 3 & \ce{^{107}Ag_3}\ce{^{79}Br_2}& $ (Ag_3Br_2)^+$\\
 &    & 482.550   & & \ce{^{107}Ag_2}\ce{^{109}Ag}\ce{^{79}Br}\ce{^{81}Br}& $ (Ag_3Br_2)^+$\\
 &    & 482.551   & & \ce{^{107}Ag_2}\ce{^{109}Ag}\ce{^{79}Br_2}& $ (Ag_3Br_2)^+$\\
485.15 &  0.043   & 484.548   & 3 & \ce{^{107}Ag_2}\ce{^{109}Ag}\ce{^{81}Br_2}& $ (Ag_3Br_2)^+$\\
&  & 484.549  & & \ce{^{107}Ag}\ce{^{109}Ag_2}\ce{^{79}Br}\ce{^{81}Br}& $ (Ag_3Br_2)^+$\\
&  & 484.551  & & \ce{^{109}Ag_3}\ce{^{79}Br_2}& $ (Ag_3Br_2)^+$\\
486.94 &  0.023   & 486.547   & 2 & \ce{^{107}Ag}\ce{^{109}Ag_2}\ce{^{79}Br_2}& $ (Ag_3Br_2)^+$\\
&  & 486.549   &  & \ce{^{109}Ag_3}\ce{^{79}Br}\ce{^{81}Br}& $ (Ag_3Br_2)^+$\\
Missing &     & 488.547   & 1 & \ce{^{109}Ag_3}\ce{^{81}Br_2}& $ (Ag_3Br_2)^+$\\
Missing &     & 664.375   & 1 & \ce{^{107}Ag_4}\ce{^{79}Br_3}& $ (Ag_4Br_3)^+$\\
Missing &     & 666.373   & 2 & \ce{^{107}Ag_4}\ce{^{79}Br_2}\ce{^{81}Br}& $ (Ag_4Br_3)^+$\\
 &     & 666.375   &  & \ce{^{107}Ag_3}\ce{^{109}Ag}\ce{^{79}Br_3}& $ (Ag_4Br_3)^+$\\
Missing &    & 668.371   & 3 & \ce{^{107}Ag_4}\ce{^{79}Br}\ce{^{81}Br_2}& $ (Ag_4Br_3)^+$\\
&    & 668.373   &  & \ce{^{107}Ag_3}\ce{^{109}Ag}\ce{^{79}Br_2}\ce{^{81}Br}& $ (Ag_4Br_3)^+$\\
&    & 668.375   &  & \ce{^{107}Ag_2} \ce{^{109}Ag_2}\ce{^{79}Br_3}& $ (Ag_4Br_3)^+$\\
669.63 &  0.016  & 670.369   & 4 & \ce{^{107}Ag_4}\ce{^{81}Br_3}& $ (Ag_4Br_3)^+$\\
&    & 670.371   &  & \ce{^{107}Ag_3}\ce{^{109}Ag}\ce{^{79}Br}\ce{^{81}Br_2}& $ (Ag_4Br_3)^+$\\
&    & 670.373   &  & \ce{^{107}Ag_2}\ce{^{109}Ag_2}\ce{^{79}Br_2}\ce{^{81}Br}& $ (Ag_4Br_3)^+$\\
&    & 670.374   &  & \ce{^{107}Ag}\ce{^{109}Ag_3}\ce{^{79}Br_3}& $ (Ag_4Br_3)^+$\\
672.08 &  0.017   & 672.369   & 4 & \ce{^{107}Ag_3}\ce{^{109}Ag}\ce{^{81}Br_3}& $ (Ag_4Br_3)^+$\\
 &     & 672.371   &  & \ce{^{107}Ag_2}\ce{^{109}Ag_2}\ce{^{79}Br}\ce{^{81}Br_2}& $ (Ag_4Br_3)^+$\\
 &     & 672.372   &  & \ce{^{107}Ag}\ce{^{109}Ag_3}\ce{^{79}Br_2}\ce{^{81}Br}& $ (Ag_4Br_3)^+$\\
 &     & 672.374   &  & \ce{^{109}Ag_4}\ce{^{79}Br_3}& $ (Ag_4Br_3)^+$\\
Missing &     & 674.369   & 3 & \ce{^{107}Ag_2} \ce{^{109}Ag_2}\ce{^{81}Br_2}& $ (Ag_4Br_3)^+$\\
&    & 674.370   &  & \ce{^{107}Ag}\ce{^{109}Ag_3}\ce{^{79}Br}\ce{^{81}Br_2}& $ (Ag_4Br_3)^+$\\
&    & 674.372   &  & \ce{^{107}Ag_4}\ce{^{79}Br_2}\ce{^{81}Br}& $ (Ag_4Br_3)^+$\\
675.68 &  0.015   & 676.368   & 2 & \ce{^{107}Ag}\ce{^{109}Ag_3}\ce{^{81}Br_3}& $ (Ag_4Br_3)^+$\\
 &    & 676.370   &  & \ce{^{107}Ag_4}\ce{^{79}Br}\ce{^{81}Br_2}& $ (Ag_4Br_3)^+$\\
Missing &     & 678.368  &  1 & \ce{^{109}Ag_4}\ce{^{81}Br_3}& $ (Ag_4Br_3)^+$\\ \hline
\end{tabular}
\caption{Comparison of the DLD mass peaks observed in the experiment with those calculated from the isotopic formula}
\label{tab:Iso}
\end{center}
\end{table}

 If the peak merge number large than one, the $\Sigma_a$ need include all the isotopic formula from this peak. From Figure ~\ref{fig:figure7}, we know that when the cluster size increase,  $\alpha_n$ will decrease. This means that the peak intensities are sensitive to the isotopic abundance composed of the cluster.  The sensitivity to the abundance composed of the cluster will decrease as the cluster size increases.  This can be explained by symmetry and probability principle in molecular cluster growth range we proposed in the previous section. Since the larger the cluster, the smaller probability for the cluster to form in the same environment. This principle can also easily be verified by the 
DLD-TOF-MS experiment. We should also able to observe all the clusters in ONLY one sample, however, compare the DLD-TOF-MS cluster spectra intensity, for $(Ag_2Br)^+$ cluster,$ I_{273nm,11hrs}<I_{269nm,5hrs}<I_{274nm,0hrs}  $; for $(Ag_3Br_2)^+$ cluster,$ I_{273nm,11hrs}>I_{269nm,5hrs}>I_{274nm,0hrs}  $. Also based on principle 3, the spectra turn around point should be the tetramer, althrough our previos theories prdict that both trimer and tetramer are possible~\cite{Zhang2000B}

\subsection{Discussion of Possible Applications of Isotopic Properties of Silver Bromide Clusters }
\begin{figure}[H]
\begin{center}
\epsfig{file=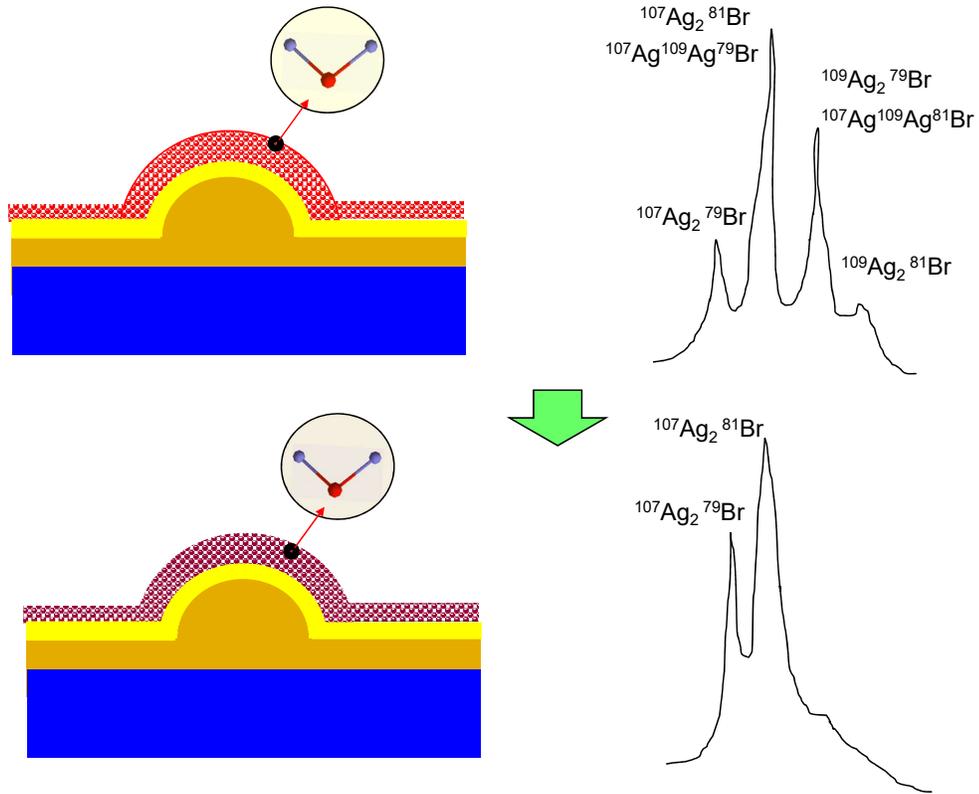,height=4.8in}
\caption{Device diagram for dark matter detection. The blue layer is substrate silicon. The dielectric layer such as $SiO_2$ sitting on top of the substrate layer. The half sphere shape is for the increase of the detection area. The yellow layer is an adhesion layer, the top red layer is the molecular silver bromide cluster layer and it can be deposited by the molecular cluster deposition tool ~\cite{Zhang2018A}}
\label{fig:figure8}
\end{center}
\end{figure}

If you add up all the matter and energy in the universe, you'd find little that is familiar. The stars and gas that astronomers see in their telescopes make up just 0.5 percent of the cosmos. Just 0.01 percent of the universe is made of elements heavier than hydrogen or helium. However  a reliable direct detection of the underlying dark-matter has remained elusive, because earlier candidates for such detections were either falsified or suffered from low signal-to-noise ratios and unphysical misalignments of dark and luminous matter. Recently  AgX emulsion grains (40 nm) for dark matter detection had been proposed by Tadaaki Tani and Tatsuhiro NakaHere~\cite{Tadaaki2016A} Here based on our previous studies, we proposed to use silver bromide isotopic molecular clusters to detect the dark matter. The principle for the detection is based on most recent finding that the isotopic atoms will have radioactive decay when interact with dark matter~\cite{Jouni}. This has been described in Figure~\ref{fig:figure8}. The top red layer are the silver bromide clusters such as $(Ag_2Br)^+$, when interact with dark matter, top red layer will change to dark red layer caused by, for example, the silver isotopic atoms ${}^{109}Ag$ decay to ${}^{107}Ag$. This will cause the mass spectra greatly changed from 5 peaks to 2 peaks as shown in Figure~\ref{fig:figure8}. It should be pointed out that the method we discussed here not restricted to silver bromide isotopic molecular cluster and dark matter detection, any environment that can cause the isotopic atom decay or have the isotopic atom abundance change can be detected by the same method. For example, environment pollution detection caused by the nuclear radiation.

\section{Conclusion}
We show that in the molecular cluster regime, why the cluster UV absorption have such wide band and it take so long time to observe the band shift? why DLD-TOF-MS spectra show an unsymmetrical cluster finger peaks? why the large clusters have much lower DLD-TOF-MS spectra intensity, all this can be systematically understood by the symmetry and probability principles of molecular growth range. The isotopic properties of silver bromide clusters and their possible application are also discussed in this paper.

\end{document}

%% file: econfmacros.tex
\def\beq{\begin{equation}}
\def\eeq#1{\label{#1}\end{equation}}
\def\eeqn{\end{equation}}

\def\beqa{\begin{eqnarray}}
\def\eeqa#1{\label{#1}\end{eqnarray}}
\def\eeqan{\end{eqnarray}}

\let\bar=\overbar

\def\Dslash{\not{\hbox{\kern-4pt $D$}}}
\def\dslash{\not{\hbox{\kern-2pt $\del$}}}

\def\msb{{\bar{\ssstyle M \kern -1pt S}}}

